\documentclass[aps,prb,twocolumn,superscriptaddress]{revtex4}

\usepackage{color}
\usepackage{bm}
\usepackage{amsmath}
\usepackage{amssymb}
\usepackage{amsthm}
\usepackage{graphicx}
\usepackage{sidecap}
\usepackage{epsfig}
\usepackage{natbib}

\begin{document}
\title{Intersubband polaritonics revisited}
\author{O. Kyriienko}
\affiliation{Science Institute, University of Iceland, Dunhagi 3,
IS-107, Reykjavik, Iceland}
\affiliation{Division of Physics and Applied Physics, Nanyang Technological University 637371, Singapore}
\author{I. A. Shelykh}
\affiliation{Science Institute, University of Iceland, Dunhagi 3,
IS-107, Reykjavik, Iceland}
\affiliation{Division of Physics and Applied Physics, Nanyang Technological University 637371, Singapore}
\email{kyriienko@ukr.net, shelykh@raunvis.hi.is}
\date{\today}

\begin{abstract}
We revisited the intersubband polaritonics -- the branch of mesoscopic physics having a huge potential for optoelectronic applications in the infrared and terahertz domains -- and found that, contrary to the general opinion, the Coulomb interactions play crucial role in the processes of light-matter coupling in the considered systems. Electron-electron and electron-hole interactions radically change the nature of the elementary excitations in these systems. We show that intersubband polaritons represent the result of the coupling of a photonic mode with collective excitations, and not non-interacting electron-hole pairs as it was supposed in the previous works on the subject.
\end{abstract}
\maketitle

\section{Introduction}
Intersubband optical transitions in a semiconductor quantum wells
(QWs) are widely used in a variety of modern optoelectronic devices
operating in a broad wavelength range spanning from mid-infrared to
terahertz. \cite{Dupont2003,Gunter,Walther,Cathabard,NatureComm}
However, their practical  realization needs high radiative quantum
efficiency. In this context the implementation of the concepts of
strong light-matter coupling is a promising tool to improve the
functionality of the devices as compare to those operating in the
weak-coupling regime. \cite{Colombelli,Jouy}

An achievement of the strong coupling is possible if the absorbing
media is placed inside a photonic cavity and coherent light-matter
coupling overcomes the dissipative processes in the system. For
intersubband transitions the experimental realization of strong
coupling regime was for the first time reported in the pioneering
work of D. Dini \textit{et al.} \cite{Dini} The elementary
excitations in this case have hybrid half-light half-matter nature
and are called intersubband polaritons. They have a number of
peculiarities distinguishing them from conventional cavity
polaritons formed by interband excitons. First, they are formed only
in TM polarization, as optical selection rules prohibit the
absorption of the TE mode in the intersubband transitions. Second, the
strength of the coupling can be an important fraction of the photon
energy, which makes possible the transition to so-called ultrastrong
coupling regime \cite{Gunter,Ultrastrong}. The light-matter coupling
constant and resulting Rabi splitting depend on the geometry of the
QW and photonic cavity and the electronic density in the lowest
energy subband \cite{DeLiberato}, which opens a possibility to tune
this parameter by application of the external gate voltage.

The broad variety of the applications of intersubband polaritonics
makes important the understanding of the nature of
intersubband polaritons. The question is: what kind of the
elementary excitation in a QW is coupled with a photonic mode and
participate in the formation of the polariton doublet? The former
can be devided into two categories: single-particle excitations
(SPE) and collective excitations, appearing from electron-electron
interactions and absent in the non-interacting system. The earlier works
devoted to theoretical description of the intersubband polaritons
neglected Coulomb interactions completely
\cite{DeLiberato,Ciuti2005,Ciuti2006,CiutiPRL} and the main
qualitative conclusion was that the formation of the polaritons is a
result of the coupling between non-interacting electron-hole pairs
and a cavity mode. Moreover, it was claimed that bosonization
approach is valid for the description of unbounded fermion pairs.
\cite{CiutiPRL}

The opinion that Coulomb effects play no substantial role in the
intersubband polaritonics seems controversial, as their important
role in photoabsorption of individual QWs (in the absence of a
photonic cavity) is an established fact, studied extensively from
both experimental and theoretical points of view.
\cite{Pinczuk1989,DasSarma1993,Vasko,Korovin} Electron-electron
interactions lead to the appearance of the collective excitation
modes such as intersubband plasmon (ISP) which under certain
conditions can give a dominant impact to the optical response.
\cite{Raikh,Marmorkos1993,Nikonov,Li2003,Pinczuk1989,DasSarma1993,Ryan,Vasko}

The effects of many-particle interactions were revealed in the case of ultra-strongly coupled intersubband transition in 0D "polaritonic dots". \cite{Ultrastrong} The emergent polariton gap in the case of ultra-strong coupling was explained using phenomenological model which includes depolarization shift. However, the consequences of Coulomb interaction in broad concentration range were not studied. Thus, the role of many-body correction for ordinary single quantum well remains uninvestigated.

In the current paper we address this question in details. We propose a semi-analytical way of the description of the coupling between
intersubband excitations (single-particle or collective) to a cavity
mode in terms of Feynman diagrams corresponding to the different
physical processes in the system. This makes our calculations
transparent and allows a simple qualitative interpretation of the
role of many body interactions in the considered system. Moreover,
since we sum up infinite series of the diagrams, our treatment is
non-perturbative and includes all orders of the interaction. As
well, it treats the resonant and anti-resonant terms in light-matter
coupling Hamiltonian on equal footing and thus allows the
description of the phenomenon of ultra-strong coupling as well. We
show theoretically that Coulomb interactions play important role in the intersubband
polaritonics, especially for high electron concentrations necessary
for the achievement of strong coupling regime. We claim that in this
case the intersubband polariton is formed due to the coupling of the
collective excitation known as intersubband plasmon (ISP) with the
cavity mode. In the opposite case of small electron concentrations
we show that excitonic corrections become crucial.

\section{The model}
\begin{figure}
\includegraphics[width=1.0\linewidth]{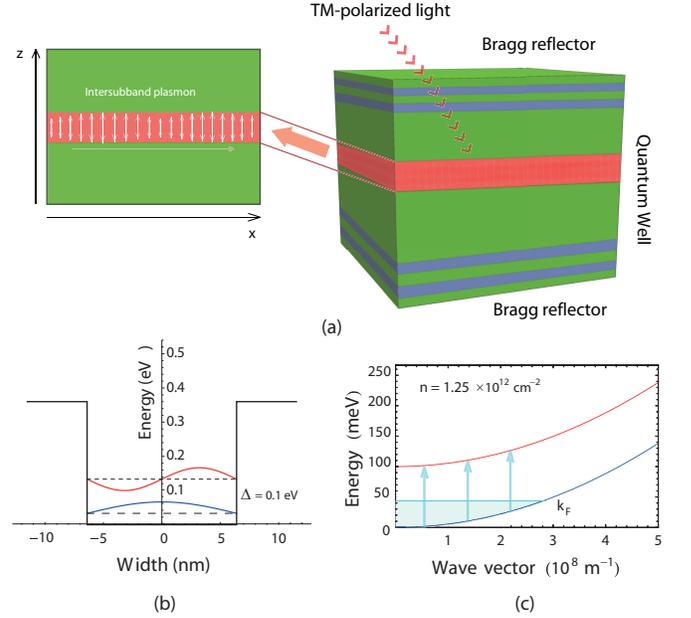}
\caption{(Color online) Geometry of the system. (a), GaAs quantum well (QW) placed into the microcavity created by distributed Bragg reflectors or air. Using the total internal reflection of light, TM-polarized beam travels through the structure exciting the excitations between upper and lower subbands. The current figure shows the creation of intersubband plasmon -- charge-density excitation. (b), Sketch of the QW of width $L=12.8$ $nm$ with plotted wave functions for fundamental and upper subbands. The separation energy between levels is ($\Delta=100$ meV). (c), Dispersions of electrons for two subbands with Fermi level energy for electron concentration $n=1.25 \times 10^{12}$ $cm^{-2}$. The single-particle excitations are schematically described as blue arrows. The $k_{F}$ label denotes to the Fermi wave vector.}
\label{Fig1}
\end{figure}

We consider a system with GaAs/AlGaAs quantum well (QW) embedded
into microcavity in the configuration usually used for intersubband
transitions with TM polarized light (Fig. \ref{Fig1}(a,b)). As our
goal here is to present qualitative results and not to perform a fit
of any experimental data, we do not consider the case where the
photonic cavity contains several QWs which is often used in order to
obtain larger values of Rabi splitting and necessary for the
achievement of the ultra-strong coupling.\cite{Gunter,Ultrastrong} Our method, however, can be easily generalized for this configuration as well. We choose a doping of a single QW in such a way that only the lowest subband is filled by
the electrons at $T=0$ while upper subbands remain empty (Fig.
\ref{Fig1}(c)). Later on we concentrate on $T=0$ case only.

\begin{figure}
\includegraphics[width=1.0\linewidth]{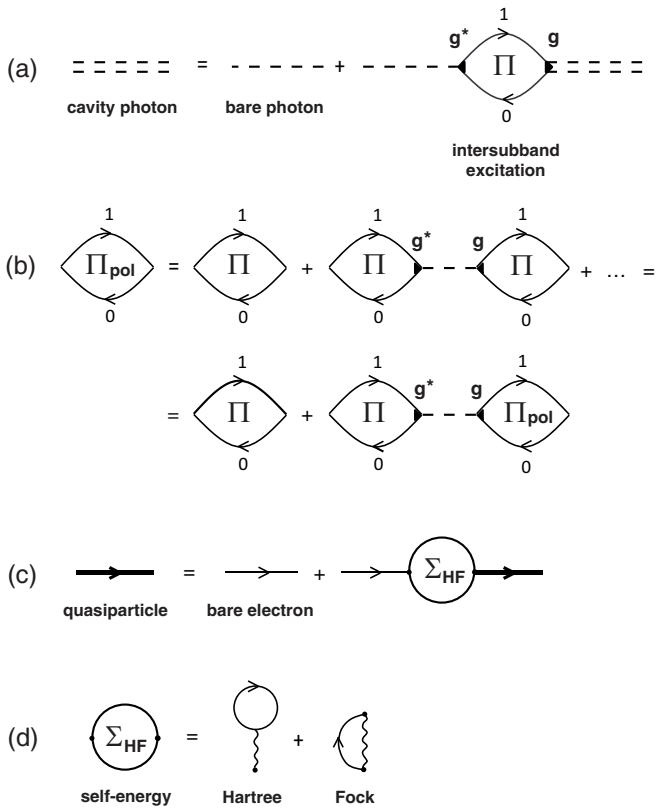}
\caption{Diagrammatic representation of intersubband polaritons. (a), The Dyson equation for microcavity photon interacting with intersubband quasiparticles $\Pi$ leading to the formation of polariton. The label $g$ corresponds to quasiparticle-photon interaction constant, indices 1 and 0 denote to excited and fundamental subband, respectively. (b), Series for absorption by intersubband quasiparticle coupled to the cavity in the general form with accounting all many-body effects. (c), The Dyson equation for self-energy corrections of the Green function of the electron. (d), The Hartree-Fock self-energy which consists of direct (Hartree) and exchange (Fock) diagrams.}
\label{Fig2}
\end{figure}

When quantum well is embedded into a microcavity, the photons
interact continuously with electrons moving them from fundamental
subband to the first excited subband, thus creating electrons and
holes which can interact with each other. This electron-hole pairs
can then again disappear, re-emitting a photon. In diagrammatic
language such process can be described by a polarization bubble
$\Pi$ (Fig. \ref{Fig2}). The photon in a cavity thus becomes
"dressed" by such bubbles, and its Green function $G$ can be
described as a sum of the terms containing one, two, three
\textit{etc} bubbles as it is shown in Fig. \ref{Fig2}(a).
The summation up to the infinite order gives the Dyson equation, whose solution yields
\begin{equation}
G=\frac{G_{0}}{1-g^{2}G_{0}\Pi },
\label{G}
\end{equation}
where $\Pi$ denotes to the full polarization operator described by
the intersubband bubble containing all possible Coulomb
interactions. In this expression $G_0$ is a bare photon Green
function,
\begin{equation}
G_{0}(\omega,q)=\frac{2\hbar\omega_{0}(q)}{\hbar^{2}\omega^{2}-\hbar^{2}\omega_{0}^{2}(q)+2i\Gamma\omega_{0}(q)},
\label{G0}
\end{equation}
where $\omega_{0}(q)$ describes the cavity mode dispersion and $\Gamma$ is the broadening of the photonic mode due to the finite lifetime (taken to be $\approx 10 ps$) and $g$ is a matrix element of electron-photon interaction which reads \cite{DeLiberato}
\begin{equation}
g(q)=\sqrt{\frac{\Delta\cdot d_{10}^{2}}{\hbar^{2} \epsilon \epsilon_{0} L_{cav}A\omega_{0}(q)}\frac{q^{2}}{(\pi/L_{cav})^{2}+q^{2}}},
\label{g}
\end{equation}
where $L_{cav}$ is cavity length, $\Delta$ is separation energy between levels, $\epsilon_{0}$ and $\epsilon$ are vacuum permittivity and relative material dielectric constant, respectively, $d_{10}$ stands for the dipole matrix element of the transition and $A$ is an area of the sample.

The spectrum of elementary excitations in such system is determined by the poles of $G(\omega,q)$ and can be found by solving a transcendental
equation
\begin{equation}
1-g^{2}G_{0}\Pi = 0.
\label{Poles}
\end{equation}

The coefficient of the photoabsorption, $\alpha(q,\omega)$, is
proportional to the imaginary part of the full polarization operator
of intersubband polariton accounting for multiple re-emissions and
re-absorptions of the cavity photon $\Pi_{pol}(q,\omega)$ can be written as
\begin{equation}
\alpha(q,\omega)\sim\omega \Im\Pi_{pol}(q,\omega).
\label{Absorb}
\end{equation}
The diagrammatic representation of the equation for $\Pi_{pol}(q,\omega)$ is shown in Fig. \ref{Fig2}(b) and yields
\begin{equation}
\Pi_{pol}=\frac{\Pi}{1-g^{2}G_{0}\Pi }.
\label{Pipol}
\end{equation}

As it is seen from the Eqs. (\ref{Poles})-(\ref{Pipol}), all
properties of intersubband polaritons can be determined if the
expression of the polarization operator of an individual QW
$\Pi(q,\omega)$ accounting for Coulomb interactions is known. In
general, the calculation of this quantity is a tricky task which can
be performed only in some particular cases, which we are now going
to consider.

\section{Non-interacting case and Hartree-Fock approximation}

This case has a methodological interest and represents a test for
our approach, allowing to compare the results it gives with those
obtained earlier in the Refs.
[\onlinecite{DeLiberato,Ciuti2005,Ciuti2006,CiutiPRL}] by using the
bosonisation scheme. For non-interacting particles the calculation
of the polarization operator $\Pi_{0}$ represented by a single
bubble without Coulomb interactions is straightforward and gives
\cite{Raikh}
\begin{equation}
\Pi_{0}(\omega,q)=2\int\frac{d\mathbf{k}}{(2\pi)^{2}}\frac{n_{0\mathbf{k}}}{\hbar\omega+E_{\mathbf{k}}^{(0)}-E_{\mathbf{k+q}}^{(1)}+i\gamma},
\label{P01}
\end{equation}
where $E_{\mathbf{k+q}}^{(1)}=\Delta+\hbar^{2}(\mathbf{k}+\mathbf{q})^{2}/2m$ and $E_{\mathbf{k}}^{(0)}=\hbar^{2}k^{2}/2m$ are dispersions of electrons in the excited and fundamental subbands, respectively, and $n_{0\mathbf{k}}$ is the Fermi distribution in the fundamental subband which can be replaced by a step-like function at $T=0$. The quantity $\gamma$ represents a non-radiative broadening of SPE. This integral can be calculated analytically for non-zero photon momentum when imaginary part of polarization operator appears naturally (see Appendix A).
The use of Eqs. (\ref{Poles})-(\ref{Pipol}) allows the determination
of the dispersions of the intersubband polaritons and
photoabsorption of the system. It is instructive to consider the
case $\gamma\rightarrow 0$ and assume that the transferred momentum
of photon $q$ is small as compare to the Fermi momentum of the
electron gas $k_F$. In this case, the equation for the energies of
the polariton modes (\ref{Poles}) reads
\begin{equation}
(\hbar\omega-\hbar\omega_{0}+i\Gamma)(\hbar\omega+\hbar\omega_{0}-i\Gamma)(\hbar\omega-\Delta)=2\omega_0nAg^{2}(q),
\label{disp}
\end{equation}
where $n$ denotes electron density in the fundamental subband and we define $N=nA$ as electron occupation number.
If we are outside the ultrastrong coupling regime ($\omega_0\gg nAg^{2}(q)$) this reduces to
\begin{equation}
(\hbar\omega-\hbar\omega_{0}+i\Gamma)(\hbar\omega-\Delta)=nAg^{2}(q),
\label{disp}
\end{equation}
which is nothing but the equation for two coupled harmonic
oscillators corresponding to a cavity mode and single-particle
excitations in the QW. The dispersions of intersubband polaritons
deduced from this equation coincide with those obtained in the
earlier works
[\onlinecite{DeLiberato,Ciuti2005,Ciuti2006,CiutiPRL}].

One should note that in the mean field (Hartree-Fock) approximation
the electron-electron corrections can be easily introduced into
consideration without substantial modification of the formalism. No
new collective excitations appear in this approach and the results
remain qualitatively the same as for the non-interacting case. We
thus consider both situations in the same section.

In diagrammatic representation the Hartree-Fock approximation
results into renormalization of electron Green functions which can
be described by the Dyson equation shown in Fig. \ref{Fig2}(c). Only
the first order diagrams corresponding to the direct Hartree term
and exchange Fock term (Fig. \ref{Fig2}(d)) are retained in the
expression for the self-energy in this approach. The Hartree term
diverges in the limit $A\rightarrow\infty$ but is compensated by the
interaction of the electrons with positive background \cite{Nikonov}
and the Fock exchange self-energy correction can be written in the
form
\begin{equation}
\Sigma_{HF}^{(i)}=-\sum_{\mathbf{k_{1}}}V_{i00i}(|\mathbf{k}-\mathbf{k_{1}}|)n_\mathbf{k_{1}}\,,
\label{SigmaHF}
\end{equation}
where $i=0,1$ and the index 0 corresponds to the fundamental subband
and index 1 to the first excited subband, $n^{i}_\mathbf{k_{1}}$ denotes
the Fermi distribution in the fundamental subband and sign "$-$" shows
that the exchange interaction decreases the energy. $V_{ijkl}(q)$
denotes a matrix element of the Coulomb interaction and reads
\begin{equation}
V_{ijkl}(q)=\frac{e^{2}}{2\epsilon\epsilon_{0}Aq}\int dz dz^{\prime} \phi_{i}(z)\phi_{j}(z^{\prime})\phi_{k}(z^{\prime})\phi_{l}(z)e^{-q|z-z^{\prime}|}
\label{Vq}
\end{equation}
with indices $i,j,k$ and $l$ corresponding to the initial and final
subbands from which particles interact and $\phi_i(z)$ being the
envelope wave functions in the direction of the structure growth
axis.\cite{Raikh,Koch}

The calculation of the polarization operator accounting for the
Hartree-Fock corrections can be done by substituting the energies of
bare electrons in the fundamental and first subbands by their
renormalized values, calculated as
\begin{eqnarray}
\widetilde{E}^{(i)}(\textbf{k})=E^{(i)}(\textbf{k})+\Sigma_{HF}^{(i)}(\textbf{k})
\end{eqnarray}
This renormalization has the following consequences. First, the
correction for populated fundamental subband is greater then for the
empty upper subband leading to the alteration of the transition
energy $\Delta$. Accounting for the negative sign of exchange
correction, one concludes that the effective gap
$\widetilde{\Delta}$ is increased. Second, in general the
self-energy is a function of momentum which leads to the
non-parabolicity of the renormalized dispersions and contributes to
the broadening of the absorption line. For intersubband polaritons
the first effect shifts the anticrossing point in the region of
larger momenta (several meV in the geometry we consider) and second
leads to the slight decrease of the observed Rabi splitting.

\section{Intersubband plasmon-polariton}

\begin{figure}
\includegraphics[width=1.0\linewidth]{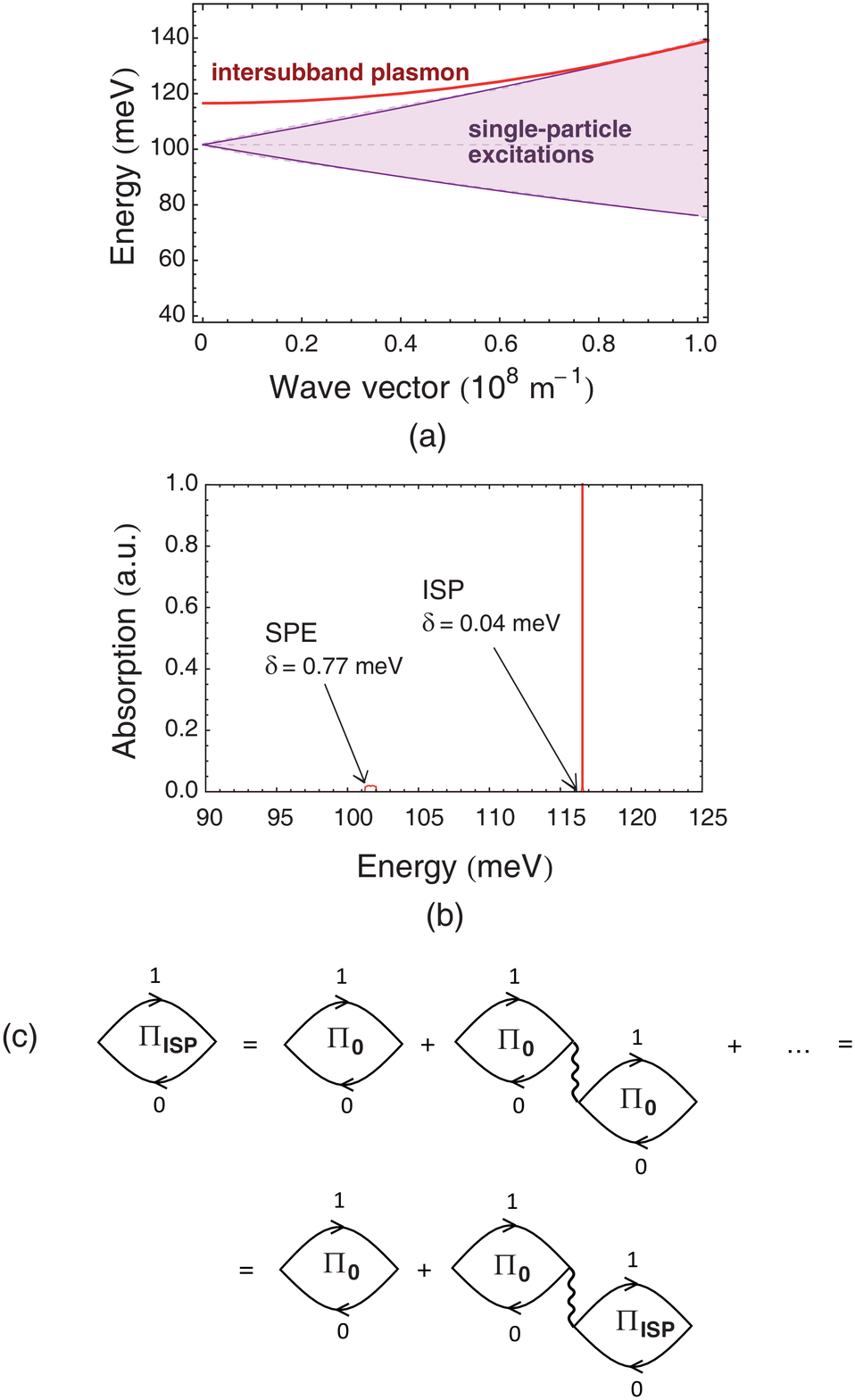}
\caption{(Color online) Intersubband plasmon. (a), The dispersion of intersubband plasmon (ISP, red line) and single-particle excitations spectrum
(SPE, violet parabolas) plotted for the long range of momentum. For
the considered concentration  $n=1.25 \times 10^{12}$ $cm^{-2}$
ISP dominates over SPE in the small wave vectors range. (b),
Absorption spectrum of intersubband quasiparticles plotted for
momentum $q=10^{6}$ $m^{-1}$. While the intersubband plasmon peak is
sharp ($\delta=0.05$ $meV$) and high, the SPE continuum is much
broader ($\delta=0.77$ $meV$) and has smaller oscillator strength.
(c), Random Phase Approximation (RPA) series for intersubband
electron-hole pairs interacting by direct Coulomb matrix element
$V_{1010}$. These diagrams correspond to the formation of
intersubband plasmon.} \label{Fig3}
\end{figure}

In general, the electron-electron interactions can not be neglected and
$\Pi\neq\Pi_0$ and their account is a non-trivial task. However, the certain types of diagrams appear in different electron density limits and govern quasiparticle dynamics. For instance, in high electron concentration case ($n\sim10^{12}~cm^{-2}$) the polarization operator can be estimated using Random Phase
Approximation (RPA) whose diagrammatic representation is shown in
Fig. \ref{Fig3}(c). \cite{DasSarma1993,Ryan} In this regime the
system demonstrates an appearance of intersubband plasmon (ISP) ---
charge-density collective excitation arising from intersubband
transitions.

The summation of the infinite series of the diagrams represented in Fig. \ref{Fig3}(c) allows us to obtain a general expression for the polarization operator corresponding to ISP
\begin{equation}
\Pi_{ISP}=\frac{\Pi_{0}}{1-V_{1010}\Pi_{0}}.
\label{Pi}
\end{equation}

Determination of the $\Pi_{ISP}$ poles in the $q=0$ limit shows that the Coulomb interaction increases the bare transition energy by the value of depolarization shift and dispersion relation is $E_{ISP}=\widetilde{\Delta}+NV_{1010}$. The absorption spectrum of intersubband plasmon can be obtained by calculation of imaginary part of polarization operator $\alpha_{ISP}\sim\omega\Im\Pi_{ISP}$, where
\begin{equation}
\Im\Pi_{ISP}=\frac{\Im\Pi_{0}(\omega,q)}{(1-V_{1010}\Re\Pi_{0}(\omega,q))^{2}+(V_{1010}\Im\Pi_{0}(\omega,q))^{2}},
\label{ImPisp}
\end{equation}
where real and imaginary part of bare intersubband bubble are given in Appendix A by Eqs. (\ref{ReFin}) and (\ref{ImFin}), respectively. One can see that appearance of depolarization term in denominator leads to the redistribution of oscillator strength comparing to the bare case $\Im\Pi_{0}(\omega,q)$.

Therefore, the absorption spectrum of intersubband plasmon polariton can be calculated by straightforward evaluation of Dyson equation written in Fig. \ref{Fig2}(b) (Eq. \ref{Pipol}) and yields
\begin{equation}
\Im\Pi_{pol}=\frac{\Im\Pi_{ISP}+g^{2}\Im G_{0}|\Pi_{ISP}|^{2}}{1+2g^{2}(\Im G_{0}\Im\Pi_{ISP}-\Re G_{0}\Re\Pi_{ISP})+g^{4}|G_{0}|^{2}|\Pi_{ISP}|^{2}},
\label{ImPisppol}
\end{equation}
where we defined absolute values $|\Pi_{ISP}|^{2}=(\Re\Pi_{ISP})^{2}+(\Im\Pi_{ISP})^{2}$ and $|G_{0}|^{2}=(\Re G_{0})^{2}+(\Im G_{0})^{2}$ and photon Green function is given by Eq. (\ref{G0}). This equation allows us to plot the intensity of intersubband plasmon polariton absorption as a function of frequency and cavity photon momentum. 

The dispersion of intersubband plasmon is shown in Fig. \ref{Fig3}(a). The absorption by intersubband plasmon is given by imaginary part of polarization $\Pi_{ISP}$ (Fig. \ref{Fig3}(b), Eq. \ref{ImPisp}). One sees that the system still has an absorption peak corresponding to the single-particle excitations. However, another peak corresponding to the absorption of ISP appears. This peak is blueshifted by a value of depolarization shift $NV_{1010}(q)$. It is much more intensive and narrow that the peak corresponding to SPE. Therefore, it is natural to suppose that after placing of the QW in a photonic cavity this peak will give main contribution to the formation of intersubband polariton.

\begin{figure}
\includegraphics[width=1.0\linewidth]{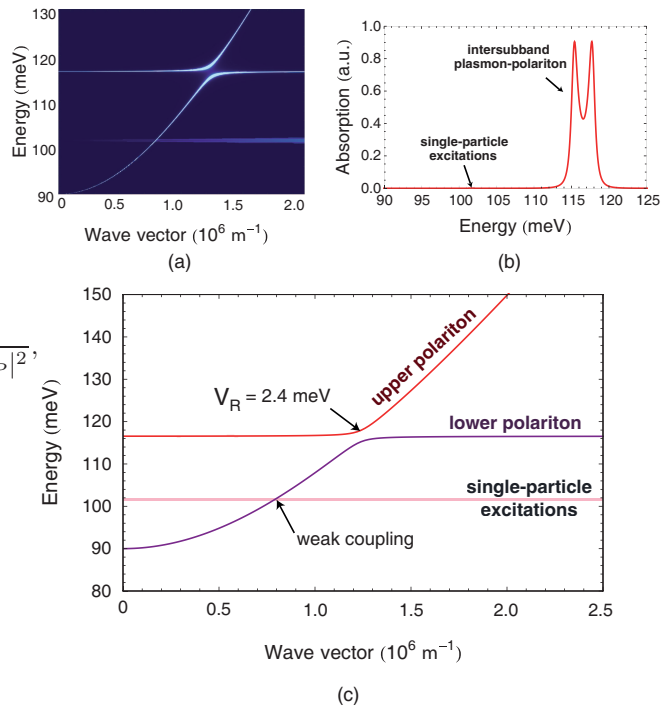}
\caption{(Color online) Intersubband plasmon polariton. (a), Density plot of the
intersubband plasmon polariton spectrum showing the dispersion of
excitations and absorption (color intensity) in the system. The
detuning of cavity mode is 10 meV. (b), Absorption spectrum of the
system plotted for wave vector $q=1.25 \times 10^{6}$ $m^{-1}$ where
the anticrossing point for ISP-photon exists. The two peaks
corresponding to upper and lower plasmon polaritons are clearly
observed, while single-particle excitations are suppressed. (c),
Dispersions of intersubband plasmon polariton modes (red and violet
lines) in strong coupling regime with corresponding anti-crossing
and Rabi frequency $V_{R}=2.4$ $meV$. The SPE dispersion (pink line)
is weakly coupled to the cavity mode.} \label{Fig4}
\end{figure}

This conclusion is supported by the results of calculations shown in Fig. \ref{Fig4}. The dispersion of the elementary excitations of the hybrid QW-cavity system was calculated using Eq. (\ref{Poles}), where the Hartree-Fock corrections were accounted for in $\Pi_0$. The corresponding spectrum of intersubband excitations in the semiconductor microcavity contains three branches (Fig. \ref{Fig4}(c), Eq. \ref{ImPisppol}). Two of them corresponding to the cavity photons and ISP reveal anticrossing and give birth to the intersubband polariton modes, which in this case can be called more correctly intersubband plasmon-polaritons. The third mode denotes to the single-particle excitations. The corresponding dispersion line crosses the dispersion of the cavity photon, which means that for SPE the weak coupling regime is realized. In the absorption spectra shown in Fig. \ref{Fig4}(b) three peaks appear. Two of them corresponding to intersubband plasmon-polaritons are very pronounced and the third one corresponding to coupling with single-particle excitations is very weak but still observable (Fig. \ref{Fig4}(a)).

\section{Excitonic effects}

\begin{figure}
\includegraphics[width=1.0\linewidth]{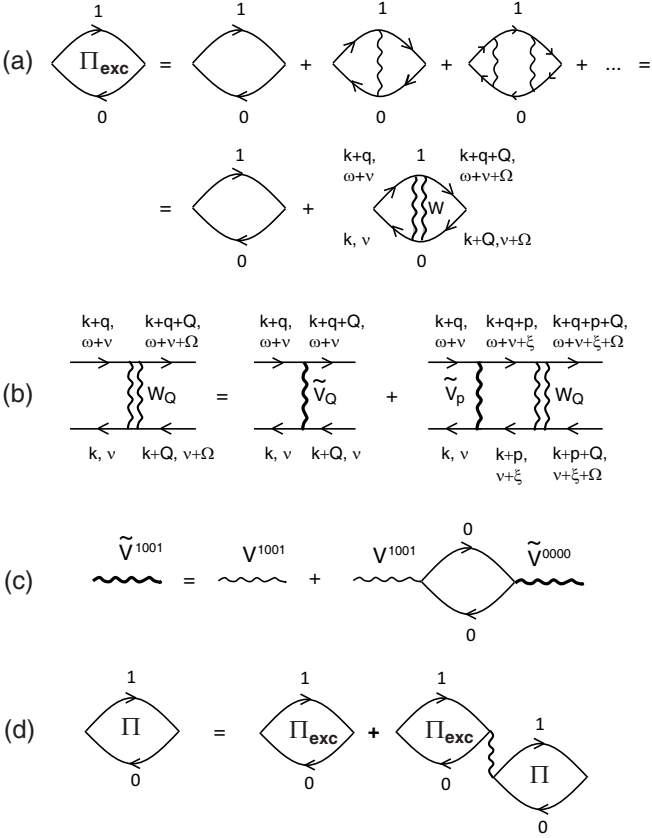}
\caption{Diagrammatic representation of intersubband exciton. (a), The ladder series for the intersubband excitation leading to the creation of intersubband exciton. The wavy line corresponds to $V_{1001}$ Coulomb interaction and double wavy line denotes to effective interaction between particles. (b), Two-particle integral equation for effective interaction. $q$ and $\omega$ are transferred momentum and energy of the photon. (c), Screening of the $V_{1001}$ electron-hole interaction by carriers in the fundamental subband. The screening by $\Pi_{10}$ processes is suppressed since $V_{1000}$ is zero. (d), The general Dyson equation which describes the situation of mixed excitonic and plasmonic effects leading to the formation of hybrid collective modes.}
\label{Fig5}
\end{figure}

In the previous section we investigated the case of high electron
concentration in QW where plasmonic effects dominate and RPA can be
successfully used. However, considering of the opposite limit of low concentrations can be also informative. In this case plasmonic corrections play minor role and excitonic effects corresponding to the interaction between the photoexcited electron in the first subband with a hole
in fundamental subband become dominant. In the diagrammatic language
they can be described by ladder diagrams shown in Fig.
\ref{Fig5}(a). \cite{Combescot,Mahan} Summation of the ladder
diagrams up to an infinite order gives birth to the formation of the
attraction between the electron and hole, and can give a peak of
photoabsorption lying below the continuum of SPE (contrary to the
the peak corresponding to ISP).

The polarization operator for intersubband exciton within the ladder approximation reads
\begin{equation}
\Pi_{exc}(\mathbf{q},\omega)=\Pi_{0}(\mathbf{q},\omega)+\Pi_{W}(\mathbf{q},\omega),
\label{Pexc0}
\end{equation}
where $\Pi_{0}$ is bare intersubband bubble and $\Pi_{W}$ denotes the intersubband bubble with accounting of effective interaction (Fig. \ref{Fig5}(a), second diagram in second line). This renormalized polarization operator can be written as
\begin{align}
&\Pi_{W}(\mathbf{q},\omega)=2\int \frac{d\mathbf{Q}}{(2\pi)^{2}}\frac{d\mathbf{k}}{(2\pi)^{2}}\frac{d\nu}{2\pi}\frac{d\Omega}{2\pi} G_{1}(\mathbf{k}+\mathbf{q},\omega +\nu) \\ \notag &G_{1}(\mathbf{k}+\mathbf{q}+\mathbf{Q},\omega +\nu +\Omega)G_{0}(\mathbf{k},\nu)G_{0}(\mathbf{k}+\mathbf{Q},\nu + \Omega)W(Q,\omega),
\end{align}
where $G_{1}(\mathbf{k},\nu)$ and $G_{0}(\mathbf{k},\nu)$ are Green functions of electron in upper subband and hole in the lower subband, respectively.

The determination of $\Pi_{W}$ requires the calculation of effective electron-hole interaction $W(Q,\omega)$, which mathematically can be deduced from the integral Bethe-Salpeter equation written in diagrammatic representation in Fig. \ref{Fig5}(b) and which reads \cite{Combescot}
\begin{align}
\label{eff_int}
&W(\mathbf{Q},\mathbf{k},\mathbf{q},\omega)=\widetilde{V}_{1001}(\mathbf{Q})+\int \frac{d\mathbf{p}}{(2\pi)^{2}}\frac{d\xi}{2\pi}\widetilde{V}_{1001}(p) \\ \nonumber
&G_{1}(\mathbf{k}+\mathbf{q}+\mathbf{p},\omega +\nu +\xi)G_{0}(\mathbf{k}+\mathbf{p},\nu +\xi)W(\mathbf{Q},\mathbf{p},\mathbf{q},\omega,\mathbf{q}),
\end{align}
where $\widetilde{V}_{1001}(Q)$ denotes an electron-hole 2D potential screened by the carriers in the fundamental subband (Fig. \ref{Fig5}(c)). Using the static intraband polarization operator which does not depend on carrier concentration in 2D case,\cite{Koch,Portnoi} one can write screened interaction
\begin{equation}
\widetilde{V}^{1001}_{Q}=-|V^{1001}_{Q}|\Big(1- \frac{\kappa}{Q+\kappa} \Big)=-\frac{e^{2}}{2\epsilon\epsilon_{0}A(Q+\kappa)},
\label{Vs}
\end{equation}
where $\kappa=\frac{me^{2}}{2\pi\epsilon\epsilon_{0}\hbar^{2}}$ is the screening wave vector and for our structure can be estimated as $\kappa=2.4\times 10^{8} m^{-1}$. In further discussion we skip the indices for exchange matrix element implying $\widetilde{V}_{Q}\equiv \widetilde{V}_{Q}^{1001}$. The Eq. (\ref{Vs}) implies that while screening by carriers in a lower subband reduces the strength of the interaction, it does not change its sign and resulted potential is still attractive.

The solution of the Bethe-Salpeter equation for 4-point vertex in a general form is a complicated task. However, we are interested in the effective interaction $W_{Q}$ (double wiggly line in Fig. \ref{Fig5}(b)). Therefore, the corresponding problem can be solved by T-matrix approach usually used for description of superconductivity.\cite{Pines,Galitskii} Writing the "in" and "out" momenta one should account that electron and a hole are created simultaneously by absorption of cavity photon with momentum $\mathbf{q}$ and energy $\omega$. In the limit of small photon momentum $q\rightarrow 0$ it is possible to solve Eq. (\ref{eff_int}) analytically where integration over intermediate variables $\xi$ and $\mathbf{p}$ becomes trivial. The contour integration on $\xi$ gives intersubband polarization operator $\Pi_{0}=N/(\hbar\omega-\Delta)$ and integration on $\mathbf{p}$ with accounting of screened interaction $\tilde{V}(p)$ yields
\begin{equation}
W(Q,\omega)=\frac{\widetilde{V}_{Q}}{1-V_{0}\Pi_{0}\chi}
\label{Weff}
\end{equation}
with factor $\chi=1+\frac{\kappa}{k_{F}}\ln\Big(1+\frac{\kappa}{\kappa+k_{F}}\Big)$ coming from $\mathbf{p}$-integration and
\begin{equation}
V_{0}=-\frac{e^2}{2\epsilon\epsilon_{0}A k_{F}}
\end{equation}
being constant Coulomb interaction for Fermi wave vector.
The poles of effective interaction give the dispersion of elementary excitations
\begin{equation}
\hbar\omega = \Delta + \chi NV_{0} = \Delta - N W_{exc},
\end{equation}
where we defined $W_{exc}=\chi|V_{0}|$. This corresponds to the lowered quasiparticle energy as compared to bare ISB case and reveals the formation of exciton-like state.

The calculation of renormalized polarization operator for intersubband exciton $\Pi_{W}$ using effective interaction (\ref{Weff}) is straightforward. It requires integration on intermediate variables $\Omega$, $\nu$, $\mathbf{k}$ and $\mathbf{Q}$ and in $q\rightarrow 0$ limit yields
\begin{align*}
&\Pi_{W}(\omega)=2\int \frac{d\mathbf{Q}}{(2\pi)^{2}}\frac{d\mathbf{k}}{(2\pi)^{2}}\frac{d\nu}{2\pi}\frac{d\Omega}{2\pi} G_{1}(\mathbf{k},\omega +\nu) \\ & G_{1}(\mathbf{k}+\mathbf{Q},\omega +\nu +\Omega)G_{0}(\mathbf{k},\nu)G_{0}(\mathbf{k}+\mathbf{Q},\nu + \Omega)W(Q,\omega).
\end{align*}
After contour integration on energy variables this expression can be rewritten as
\begin{equation*}
\Pi_{W}(\omega)=2\int \frac{d\mathbf{Q}}{(2\pi)^{2}}\frac{d\mathbf{k}}{(2\pi)^{2}} \frac{\widetilde{V}(Q)}{(\hbar\omega - \Delta + N W_{exc})(\hbar\omega - \Delta)}
\end{equation*}
and with performing momentum integration this leads to the total polarization operator in the form
\begin{align}
\Pi_{exc}&=\frac{N}{\hbar\omega - \Delta} - \frac{N}{\hbar\omega - \Delta}W_{exc}\frac{N}{\hbar\omega - \Delta + NW_{exc}}=\\
\notag &=\frac{N}{\hbar\omega - \Delta + NW_{exc}}. 
\label{Pexc}
\end{align}
The accounting of damping allows one to obtain the absorption by the intersubband exciton. However, one should note that full determination of quasiparticle absorption with non-zero $q$ requires the numerical calculation of polarization operator $\Pi_{exc}$.  

\begin{figure}
\includegraphics[width=1.0\linewidth]{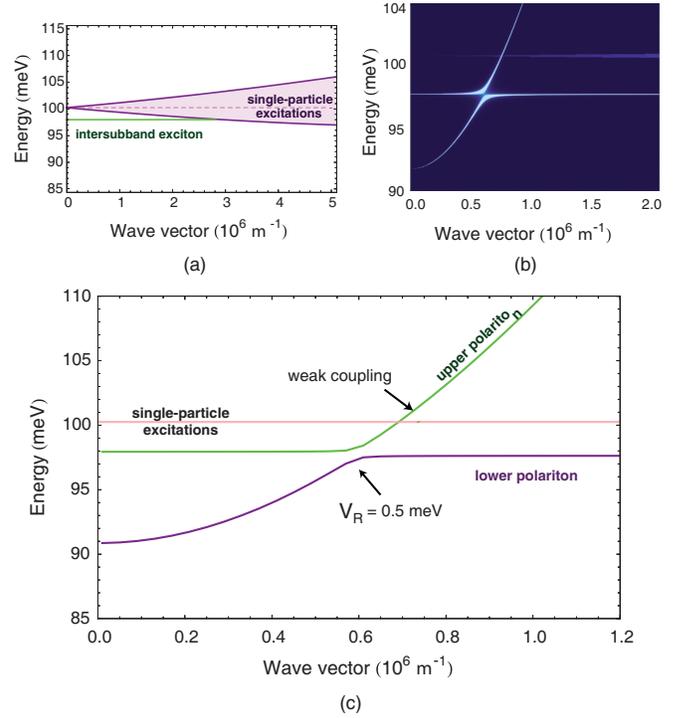}
\caption{(Color online) Intersubband exciton polariton. (a), The dispersion of
intersubband exciton (green line) which lies below the
single-particle excitations continuum (violet parabolas). The
concentration of electrons in QW is $n=10^{11}$ $cm^{-2}$. (b),
Density plot of intersubband exciton coupled to the photonic mode.
The intensity describes the absorption in the system. (c),
Dispersions of intersubband exciton polariton modes (green and
violet lines) in strong coupling regime ($V_{R}=0.5$ $meV$). The SPE
dispersion (pink line) is weakly coupled to a cavity mode.}
\label{Fig6}
\end{figure}

The calculated dispersion of intersubband exciton is shown in Fig.
\ref{Fig6}(a) for the QW with doping $n=10^{11}$ $cm^{-2}$ and resulted dispersion of the elementary excitations of the QW coupled to a cavity mode accounting for the excitonic effects are shown in Fig.
\ref{Fig6}(c). Similarly to the case of the high concentrations with
the strong ISP-photon coupling, one sees three dispersion branches.
Two of them corresponding to the excitonic and photonic modes reveal
anticrossing and form the intersubband exciton polaritons. The third
one corresponding to the single-particle excitations remains in a weak
coupling regime. In the photoabsorption spectrum shown in the Fig.
\ref{Fig6}(b) three peaks of different intensities are observed. The
excitonic effects lead to the redistribution of the oscillator
strength, which becomes small for SPE transitions and corresponding
peak is consequently very weak. In this case the Rabi splitting is
reduced due to the concentration dependence of interaction constant
and is about 0.5 meV.

It should be noted that for very low concentrations case even small intensities of the photonic pump will move most of the carriers to the upper subband. For this situation our considerations become not valid, as we supposed that the number of electronic excitations is small. However, we believe that this situation has methodological interest since is shows that strong coupling regime can be in principle achievable with collective excitations different from intersubband plasmons.

\begin{figure}
\includegraphics[width=1.0\linewidth]{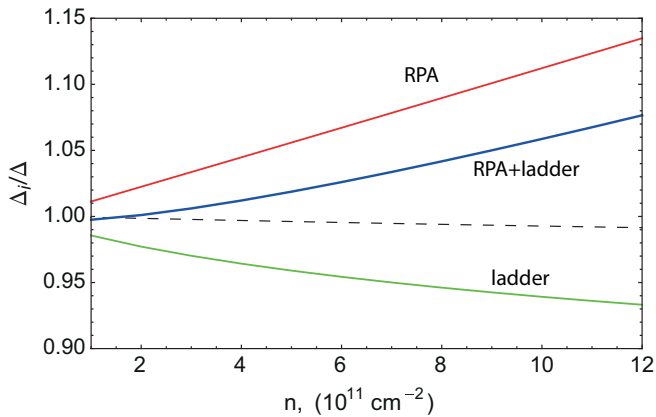}
\caption{(Color online) Concentration dependence of Coulomb interaction corrections to the intersubband energy distance. The ratio $\Delta_{i}/\Delta$ denotes different approximations for calculations, namely RPA (red line), ladder approximation (green line) and mixed ladder-RPA corrections expressed by diagrams written in Fig. \ref{Fig6}(d) (blue line). The black dashed line corresponds to exchange Hartree-Fock corrections.}
\label{Fig7}
\end{figure}

In the previous sections we studied the system with high and low concentrations which correspond to RPA and ladder approximation, respectively. In the Fig. \ref{Fig7} we present the resulted shift of intersubband transition due to accounting of different type of corrections (red line for RPA and green for ladder approximation). However, in the case of intermediate concentrations the accounting of many-electron effects becomes non-trivial. While in general it cannot be done using particular approximation, there are several ways for treatment of intermediate concentration region. Here we used the approach proposed in the Refs. [\onlinecite{Nikonov,DasSarma1993}], where both RPA and ladder corrections were accounted simultaneously. The corresponding diagrammatic series are shown in the Fig. \ref{Fig5}(d) which lead to the appearance of mixed collective exciton-plasmon modes in the system. In the Fig. \ref{Fig7} we plot the ratio of corrected energies $\Delta_{i}$ to bare ISP $\Delta$, where index $i$ stands for the type of approximation. One can see that in this general case the depolarization and excitonic corrections tend to cancel each other and resulted shift from bare transition energy is diminished (blue line). Due to the stronger concentration dependence of depolarization shift, it is predominant in the high and medium range of concentrations, while excitonic corrections become important for the $n<10^{11}$ $cm^{-2}$.

The generalization of the theory for single QW, which is presented in article, to case of $n_{QW}$ identical QWs can done by increasing the Rabi frequency $\Omega_{R}=\sqrt{n_{QW}}\Omega_{R0}$, where by $\Omega_{R0}$ we denote the Rabi frequency in single quantum well. An important feature of the theory is an appearance of the third peak corresponding to the bare ISB transition. This was recently observed in the experiment by M. Geiser \textit{et al.},\cite{Swiss} where two polaritonic modes and bare ISB mode lying between UP and LP were measured. While the influence of collective excitations is mentioned in the article, the origin of uncoupled SPE absorption mode was not explained. The authors attributed it to intersubband absorption in the bare epitaxial layer between the resonators. In the same time this peak changes the position with different doping of the QW which corresponds to influence of many-electron effects and corresponding correction of the transition frequency. 

Our theory gives qualitative explanation of observed phenomena namely that a weakly coupled bare mode is present in absorption spectrum together with polaritonic modes. However, the quantitative comparison with the experiment is complicated due to the fact that measurements were performed for parabolic quantum wells (and not rectangular QW that we considered). 
One should note that in an ideal parabolic QWs the many body electron-electron effects vanish and bare transition is not renormalized as stated in Kohn’s theorem.\cite{Kohn} Meanwhile, for the case of real wide parabolic quantum wells the contribution of depolarization and Hartree-Fock corrections usually do not cancel each other leading to the small shift from bare energy.\cite{DasSarma1993} Application of our model for this system leads to appearance of polaritonic modes based on collective intersubband plasmon excitation and red-shifted bare transition according to Hartree-Fock corrections. While general trend is correct, the exact energy position of our weakly coupled peak does not coincide with experiment. To improve it the accounting of different corrections for the case of parabolic QW has to be performed, which is formidable and cannot be done in the frame of the present paper.

\section{Conclusions}
In conclusion, we analyzed the elementary excitations arising from
the strong coupling of a photonic cavity mode with an intersubband
transition of a single QW. We have shown that contrary to the
current opinion Coulomb interactions can play crucial role in the
system and lead to the qualitative changes of the nature of
intersubband polaritons. We predict theoretically that strong
coupling of the cavity mode occurs with collective excitations,
while single-particle excitations remain in the weak coupling
regime. 
The different electron concentration regimes are considered and possible connection to experimental observation of the proposed phenomena was stated.

This work was supported by Rannis "Center of Excellence in
Polaritonics" and FP7 IRSES project "POLAPHEN". I.A.S. acknowledges
the support from COST POLATOM program. O.K. acknowledges the help of
Eimskip Foundation.

\appendix
\section{Calculation of a polarization operator for non-interacting particles}

The generic form of the electron-hole polarization operator for non-
interacting particles can be written as
\begin{equation}
i\Pi_{0}(\omega,\mathbf{q})=2\int \frac{d\mathbf{k}d\nu}{(2\pi)^4}G(\mathbf{k}+\mathbf{q},\nu + \omega)G(\mathbf{k}, \nu),
\label{P0}
\end{equation}
where $G(\mathbf{k},\nu)$ denotes the Green function of particle with momentum $\mathbf{k}$ and energy $\nu$. $\mathbf{q}$ and $\omega$ correspond to the transferred momentum and energy, respectively. For the intersubband transition case the polarization bubble describes the excitation process where electron is transferred to the upper subband while the hole is created in the lower subband. Thus, the Eq. (\ref{P0}) can be represented as
\begin{equation}
\Pi_{0}(\omega,q)=2\int\frac{d\mathbf{k}}{(2\pi)^{2}}\frac{n_{0\mathbf{k}}}{\hbar\omega+E_{\mathbf{k}}^{(0)}-E_{\mathbf{k+q}}^{(1)}+i\gamma},
\label{P01sup}
\end{equation}
where $E_{\mathbf{k}+\mathbf{q}}^{(1)}=\Delta+\hbar^{2}(\mathbf{k}+\mathbf{q})^{2}/2m$ and $E_{\mathbf{k}}^{(0)}=\hbar^{2}k^{2}/2m$ are dispersions of electrons in the excited and fundamental subbands, respectively, $\Delta$ is an energy distance between subbands and $n_{0\mathbf{k}}$ is the Fermi distribution in the fundamental subband which can be replaced by a step-like function at $T=0$. Parameter $\gamma$ describes the lifetime of the excitation.

First, let us rewrite Eq. (\ref{P01sup}) in the following form
\begin{widetext}
\begin{equation}
\Pi_{0}(\omega,q)=2\int_{0}^{k_{F}}\frac{{kdk}}{(2\pi)^{2}}\int_{0}^{2\pi}\frac{d\phi}{\hbar\omega-\Delta-\hbar^{2}q^{2}/2m-\hbar^{2}kq\cos\phi/m+i\gamma},
\label{P01rewr}
\end{equation}
\end{widetext}
where $\phi$ is an angle between two vectors $\mathbf{k}$ and $\mathbf{q}$.
The assumption that transferred momentum of photon $q$ is small yields simple integration on $\phi$ and $k$, and the real part of polarization operator is
\begin{equation}
\Re \Pi_{0}(\omega)_{q\rightarrow 0}=\frac{n A}{\hbar\omega-\Delta},
\label{ReP10long}
\end{equation}
where $n=\frac{k_{F}^{2}}{2\pi}$ is 2D density of electron gas, $A$ is QW area and we assumed $\gamma\rightarrow 0$.

Now we return to the case of non-negligible photon momentum and calculate the imaginary part of polarization operator (\ref{P01rewr}). The analysis of the denominator shows that integral on $k$ in Eq. (\ref{P01rewr}) has poles only in certain angular range [$\phi_{min},\phi_{max}$]:
\begin{equation}
\arccos\left[\frac{(\hbar\omega-\Delta-E_{q})}{\hbar^{2}k_{F}q/m}\right] < \phi < \arccos\left[\frac{-(\hbar\omega-\Delta-E_{q})}{\hbar^{2}k_{F}q/m}\right].
\label{phi}
\end{equation}
Thereby, in this region one can find the imaginary part of polarization operator using residue theorem in the limit $\gamma \rightarrow 0$:
\begin{equation}
\Im\Pi_{0}(\omega,q)_{\gamma\rightarrow 0}=\frac{Am}{4\hbar^{2}}\frac{1}{E_{q}}(\hbar\omega-\Delta-E_{q})\tan\phi|_{\phi_{min}}^{\phi_{max}}\equiv I_{\delta}(\omega,q),
\label{Im1}
\end{equation}
which describes a peak in the region where single-particle
excitations exist, bounded by parabolas
$\Delta+\hbar^{2}q^{2}/2m-\hbar^{2}k_{F}q/m$ and
$\Delta+\hbar^{2}q^{2}/2m+\hbar^{2}k_{F}q/m$.

However, to describe the realistic absorption spectral function one
needs to account for the finite non-radiative lifetime of
intersubband excitations. It can be done using the
Sokhatsky-Weierstrass theorem with non-zero $\gamma$
\begin{equation}
\int\frac{f(x)}{x+i\gamma}dx=-i\pi  \int\frac{\gamma f(x)}{\pi(x^{2}+\gamma^{2})}dx + \int\frac{x f(x)}{(x^{2}+\gamma^{2})}dx.
\label{S-W}
\end{equation}
The integral (\ref{P01rewr}) can be rewritten in the similar form
\begin{eqnarray}\label{P01SW}
&\Pi_{0}(\omega,q)=-2\int_{0}^{2\pi}\frac{{d\phi}}{(2\pi)^{2}}\int_{0}^{k_{F}}\frac{k}{Bk+C+i\gamma}dk=\\ \nonumber
&=-\frac{1}{2\pi^2}\int_{0}^{2\pi}\frac{{d\phi}}{B}\int_{0}^{k_{F}}\frac{k}{k-k_0+i\gamma}dk=\\ \nonumber
&=-\frac{1}{2\pi^2}\int_{0}^{2\pi}\frac{{d\phi}}{B}\int_{-k_0}^{k_{F}-k_0}\frac{\widetilde{k}+k_0}{\widetilde{k}+i\gamma}d\widetilde{k},
\end{eqnarray}
where $B=\hbar^{2}q\cos\phi/m$, $C=\hbar\omega-\Delta-\hbar^{2}q^{2}/2m$, $k_0=C/B$ and the $\phi$-dependence of $B$ and $k_0$ should be taken into consideration. In the latest expression the new integration variable $\widetilde{k}=k-k_0$ was used.
Thus, comparing the Eqs. (\ref{P01SW}) and (\ref{S-W}), one can calculate numerically the imaginary part of polarization operator (\ref{P01rewr}) with finite lifetime of excitations.
One sees that integral (\ref{P01SW}) can be separated into two parts. Detailed analysis shows that the first integral gives obtained previously delta-function like absorption which does not depend on $\gamma$, while the second integral $I_{Lor}$ is proportional to $\gamma$ and has Lorentzian form.
Therefore, for the sake of simplicity, in the purely analytical calculations it is possible to use absorption spectrum as the sum of equation (\ref{Im1}) and phenomenological Lorentzian broadening due to finite lifetime of excitation ($\tau \approx\mu s$) written in the form \cite{Koch}
\begin{equation}
I_{Lor}(\omega)=\Im\left[\frac{nA}{\hbar\omega-\Delta+i\gamma}\right]=\frac{-\gamma nA}{(\hbar\omega-\Delta)^{2}+\gamma^{2}}.
\label{ImP10long}
\end{equation}

Finally, the imaginary part of polarization operator which describes optical absorption by intersubband transition yields
\begin{equation}
\Im \Pi_{0}(\omega,q)=I_{\delta}(\omega,q)+I_{Lor}(\omega)
\label{ImFin}
\end{equation}

The real part of polarization operator can be found by direct integration in Eq. (\ref{P01rewr}) \cite{Raikh,Koch}
\begin{align}
\nonumber
\Re\Pi_{0}(\omega,q)=\frac{Am}{\pi \hbar^{2}}\frac{1}{2E_{q}} ( (\hbar\omega -\Delta - E_{q}) \mp \\
\mp \sqrt{(\hbar\omega - \Delta - E_{q})^{2}-4E_{F}E_{q}} ),
\label{ReFin}
\end{align}
where $"-"$ sign corresponds to the case $\hbar\omega>\Delta$ and $"+"$ for $\hbar\omega<\Delta$.

Consequently, the polarization operator $\Pi_{0}$ which stands for the simple intersubband bubble is described by the sum of real (Eq. (\ref{ReFin})) and imaginary (Eq. (\ref{ImFin})) parts. Finally, using it in the Dyson equations, the optical response and elementary excitation spectrum can be found.


\end{document}